# On the elementary theory of earthquake aftershocks

A.V. Guglielmi, O.D. Zotov

*Schmidt Institute of Physics of the Earth, Russian Academy of Sciences; Bol'shaya Gruzinskaya str., 10, bld. 1, Moscow, 123242 Russia; guglielmi@mail.ru (A.G.), ozotov@inbox.ru (O.Z.)*

**Abstract**

The elementary theory of aftershocks, being relatively simple mathematically, belongs to the basics of earthquake physics. The paper briefly outlines the concepts and ideas of the theory, provides equations for the relaxation of the source after the main shock, and emphasizes the effectiveness of the theory in the search for new approaches to the processing and analysis of aftershock observation data. The main attention is paid to the discussion of two new results obtained on the basis of elementary theory. Firstly, a two-stage relaxation mode of the earthquake source was discovered. In the first stage, called the Omori epoch, the Omori law is strictly observed. The first stage ends with a bifurcation of the source, and the second stage of relaxation begins, during which the evolution of aftershocks proceeds chaotically. Secondly, within the framework of the elementary theory of aftershocks, a fundamentally new concept of the proper time of the source was introduced.

***Keywords*:** earthquake source, deactivation coefficient, Omori law, dynamic system, inverse problem, relaxation, proper time. The source proper time has proven useful in experimental studies of seismic activity.

## 1. Introduction

Following the main shock of a tectonic earthquake, a series of aftershocks is observed, the frequency of which $n(t)$ decreases with time. It was for aftershocks that the first law of earthquake physics was established back in the century before last – the so-called Omori law, according to which the average frequency $n(t)$ decreases hyperbolically over time [6]. Here it is appropriate to point out numerous works [7–16] devoted to Omori's law. Bath's law has also been established, according to which the difference between the magnitude of the main shock and the maximum magnitude of the aftershocks exceeds 1.2 [17, 18].

The Omori and Bath laws, as well as other interesting properties of aftershocks, were established empirically without relying on theory. The search for ways to create a theory of earthquakes is being conducted by geophysicists in various directions, one of which is presented in a series of publications by A.D. Zavyalov, B.I. Klain and the authors of this paper [19–25]. We proceed from the idealized concept of the source as a dynamic system and construct a deductive theory of relaxation of the source, "cooling down" after the main shock. The main parameter of the theory is the so-called deactivation coefficient of the source $\sigma(t)$, which can be calculated based on the observation data of the frequency of aftershocks $n(t)$.

We call our theory elementary not only because it relates to the fundamentals of earthquake physics, but also because, mathematically, the theory is at a simple, initial level of development. The relationship between time and the state of the system is defined axiomatically. An equation describing the evolution of aftershocks is derived. It is important that no restrictions are imposed on the form of the function $\sigma(t)$, except for the requirement of continuity and smoothness. We will give a synoptic overview of the elementary theory in the next section of the paper.

At the same time, despite its simplicity, the elementary theory has proven effective in finding new approaches to processing and analyzing aftershocks. From the variety of results obtained on the basis of elementary theory, we will highlight and discuss in detail the two-stage relaxation mode of the source, and a conceptually new idea of the proper time of the source.

The proposed theory is phenomenological. Firstly, this means that the deactivation coefficient and other phenomenological parameters are not interpreted in general physical concepts and ideas, but are calculated on the basis of experimental data. Secondly, in constructing the theory we used the methodological techniques of transcendental phenomenology, namely, reduction and epoché, or, to put it simply, reducing the complex to the simple and refraining from a priori judgments about the nature of the relaxation of the source. In particular, and this is important to emphasize, we have abandoned, unlike Omori, Hirano, and Utsu, the selection of a fitting formula for approximating the attenuation of aftershock activity over time.

## 2. Elementary theory

The phenomenological theory of aftershocks is based on the idea of an earthquake source as a dynamic system, the state of which is characterized by the parameter $\sigma(t)$, which we call the source deactivation coefficient. Formally, the value of the deactivation coefficient is equal to

$$\sigma = -\frac{1}{n^2}\frac{dn}{dt}, \qquad (1)$$

where $n(t)$ is the frequency of aftershocks, which on average decreases over time. On the phase plane of the $(\sigma,\theta)$, where $\theta = d\sigma/dt$, the representative point of the $[\sigma(t),\theta(t)]$ moves along a certain phase trajectory, which graphically represents the relaxation process of the source after the main shock. If $\sigma = \text{const}$, then the phase trajectory degenerates into a fixed point $(\sigma,0)$. In the extended phase space $(\sigma,\theta,t)$, the degenerate phase trajectory has the form of a straight line parallel to the time axis. Degeneration occurs if and only if Omori's law [6] is strictly satisfied (for more details, see [25]).

Let us introduce an auxiliary function $g(t) = 1/n(t)$. When using the new dependent variable $g(t)$, the aftershock evolution equation has the form of the simplest linear differential equation

$$\frac{dg}{dt} = \sigma(t). \qquad (2)$$

It is equivalent to the simplest nonlinear differential equation

$$\frac{dn}{dt} + \sigma n^2 = 0. \qquad (3)$$

Although formally both evolution equations are equivalent to each other, in applications of the theory to the processing of experimental data and in the search for generalizations of the elementary theory it is often more convenient to use either (2) or (3).

The inverse problem of the source is to calculate the deactivation coefficient from the observed aftershock frequency data. The correct solution, suitable for calculating $\sigma(t)$ based on real observation data $n(t)$, has the form

$$\sigma(t) = \frac{d}{dt}\langle g(t)\rangle, \qquad (4)$$

where the angle brackets denote the optimal smoothing of the auxiliary quantity $g(t)$ [5].

The elementary theory opens up new possibilities for the processing and analysis of aftershocks and provides a basis for searching for generalizations that take into account certain aspects of the dynamics of the source after the main shock. The simplest generalization consists of adding a free term to the right-hand side of the equation (3). The inhomogeneous differential equation

$$\frac{dn}{dt} + \sigma n^2 = f(t) \tag{5}$$

formally simulates the effect of a trigger on the source. For example, $f(t) \propto \delta(t - \Delta t)$ for the round-the-world seismic echo, which is an endogenous trigger excited during the main shock of an earthquake. Here $\Delta t \sim 3^h$ is the echo delay time [4, 21].

If $f(t)$ is a random function, then equation (5) becomes a stochastic differential equation. It is appropriate to use it to take into account the influence of seismic noise at the source on the flow of aftershocks.

By adding a linear term to equation (3), we obtain the Verhulst logistic equation

$$\frac{dn}{dt} = n(\gamma - \sigma n), \tag{6}$$

where $\gamma$ is the second phenomenological parameter of the source. Over time, the aftershock frequency tends asymptotically to the frequency of tremors in the background seismicity regime, $n_\infty = \gamma / \sigma$. Using $n_\infty$ with a known value of $\sigma$, $\gamma$ can be estimated.

The need for a radical generalization arose when posing and solving the problem of an experimental study of the spatio-temporal evolution of the source [20]. It was proposed to move from the ordinary differential equation of evolution (3) to a partial differential equation

$$\frac{\partial n}{\partial t} = n(\gamma - \sigma n) + D \frac{\partial^2 n}{\partial x^2}, \tag{7}$$

known as the Kolmogorov, Petrovskii and Piskunov equation (abbreviated KPP) [26]. The choice of the KPP equation is motivated by the fact that its solutions contain stationary diffusion waves propagating with velocity $v \sim \sqrt{D\gamma}$. This property of the KPP equation makes it possible to hypothetically interpret the phenomenon of aftershock propagation at a velocity of several kilometers per hour, discovered in the experiment, as the propagation of nonlinear diffusion waves [20].

## 3. Discussion

### 3.1. Relaxation mode of the earthquake source

Omori's law within the framework of elementary theory has the form $\sigma = \text{const}$. Apparently, Omori [6], and after him Hirano [7] and Utsu [9, 10] considered the law $\sigma = \text{const}$ as holistic. Meanwhile, our rich experience in solving the inverse problem of the source indicates that the deactivation coefficient undergoes significant changes during relaxation of the source. A careful analysis of the dynamics of the deactivation coefficient, as well as the phase portraits of the source, made it possible to establish that Omori's law is not holistic. We discovered a two-stage regime of aftershock evolution.

At the first stage, the Omori law $\sigma = \text{const}$ is strictly observed. The duration of the first stage varies from case to case from several days to several months. A tendency towards an increase in duration with increasing magnitude of the main shock was noted. The magnitude of the deactivation coefficient decreases with increasing magnitude of the main shock. The first stage of evolution was called the Omori epoch. During the Omori epoch, the average frequency of aftershocks is quite predictable.

The Omori era ends with a bifurcation, after which the deactivation coefficient changes rather chaotically over time. In the second stage, the evolution of aftershocks is unpredictable. Figure 2 illustrates the two-stage evolutionary regime for the event shown in Figure 1. The duration of the Omori epoch is 100 days, as shown in Figure 2.

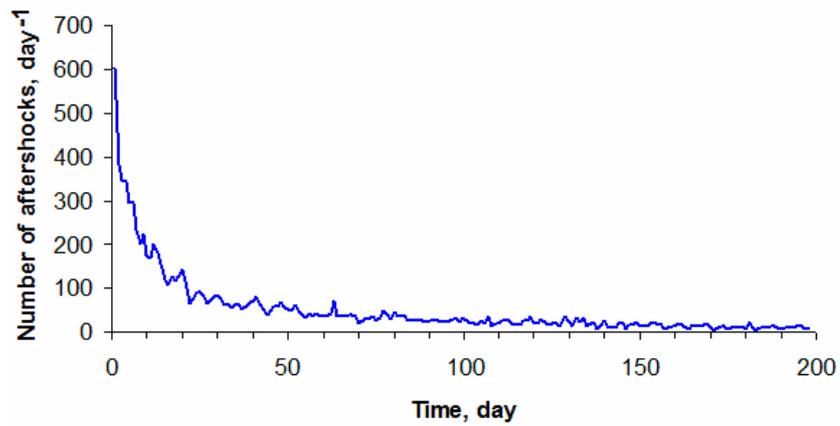

**Fig. 1**. Dependence of aftershock frequency on time. The main shock of earthquake with the magnitude of $M = 6.7$ and the hypocenter depth of 18 km occurred in Southern California on 17.01.1994. There were 9,867 aftershocks in 200 days.

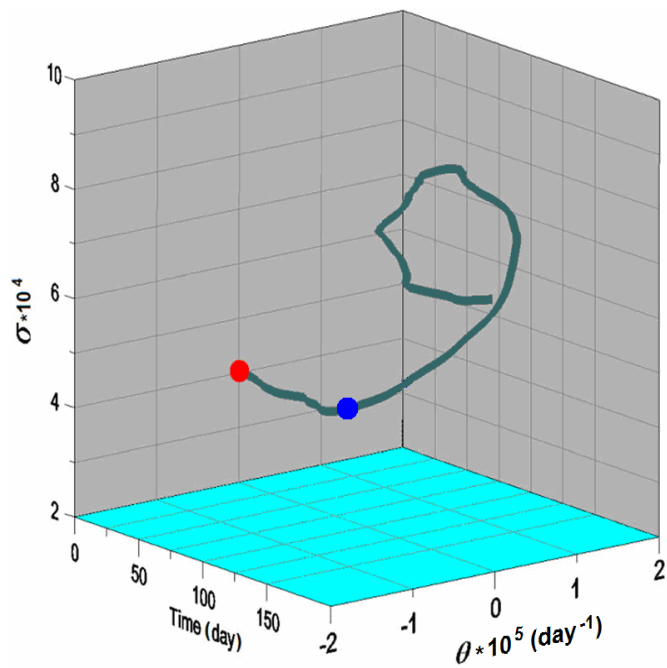

**Fig. 2**. Evolution of the earthquake source in the extended phase space. According to the data presented in Fig. 1. The red and blue dots mark the beginning and end of the Omori epoch.

In the literature it is often stated that the holistic Utsu law should be used to describe aftershocks rather than the Omori law (see, for example, [16]). But the idea of a two-stage relaxation mode of the source excludes the applicability of Utsu's holistic law. At the same time, in reality, the hyperbolic Omori law is valid, but its applicability is limited to the first stage of the evolution of the source.

### 3.2. Proper time of the earthquake source

Change of variables is an effective method for studying dynamic systems. By replacing the dependent variable, we obtained the evolution equation of the source in the form of the simplest differential equation (2). Now let's make a replacement of the independent variable $t \to \tau$, where

$$\tau = \int_0^t \sigma(t') dt'. \tag{8}$$

Let us conditionally call $\tau$ the proper time of the source. As a result of such a replacement, the nonlinear evolution equation (3) takes on a compact form

$$\frac{dn}{d\tau} + n^2 = 0. \tag{9}$$

Its general solution is equal to

$$n(t) = \frac{n_0}{1 + n_0 \tau(t)}, \tag{10}$$

where $n_0 = n(0)$ is the initial condition. We see that the aftershock frequency decreases hyperbolically with the proper time $\tau$. The property of hyperbolicity is also inherent in the classical Omori law [6], which is valid only at the first stage of evolution. A significant difference concerns the choice of independent variable. Omori used world time. We used the proper time, which, generally speaking, flows unevenly (see (8)).

The uneven flow of proper time is explained as follows. The source as a dynamic system is unstable. The rocks that make up the source are in a non-stationary state. As a result, the deactivation coefficient, which generally characterizes the functioning of the source, depends on time. By definition (8), the proper time of the source proceeds unevenly.

The idea of the proper time of the source as a continuous function of world time $\tau(t)$ is a far-reaching idealization. The idea is based on rich experience in observing discrete sequences of aftershocks. The procedure for forming the function $\tau(t)$ begins with dividing the aftershock sequence into relatively small time segments $\Delta t = t_{i+1} - t_i$, $i = 1, 2, 3,..$ of equal duration, each of which contains a sufficiently large number of aftershocks so that it is possible to $i = 1, 2, 3,..$ calculate the "instantaneous" frequency $n(t_i)$, related, for example, to the middle of each segment.

Here we cannot dwell on the methodologically complex issues associated with the choice of the value of $\Delta t$. We only note that in each interval $\Delta t$ the events have a Poisson distribution, and the sequence of average waiting times for the next aftershock $T_i$ forms an arithmetic progression in the Omori epoch. The general term of the progression is equal to

$$T_i = T_1 + (i-1)\delta T, \qquad (11)$$

where $\delta T = \sigma \Delta t$ is the step of the progression. In other words, at each moment $t_i$ the waiting time is equal to the waiting time at the previous moment $t_{i-1}$, increased by the same number: $T_i = T_{i-1} + \delta T$, $i \geq 2$. The sequence $T_i$ is monotonically increasing for $\sigma > 0$.

Next, we plot all the points on the coordinate plane $(i, T_i)$, approximate them with a smooth function $g(t)$, using which we calculate the deactivation coefficient $\sigma(t)$ and the proper time of the source $\tau(t)$. Thus, idealization consists of the transition from a finite set of numbers to a continuous smooth function. The radical nature of the transition consists in replacing the countable set that we actually observe with a set of the power of the continuum.

Our idealization was successful. It made it possible to formulate an elementary theory of aftershocks, which allowed a new approach to the analysis of aftershocks and the discovery of a number of previously unknown properties of the earthquake source. However, the new methodology is only directly applicable to the processing and analysis of aftershocks. Meanwhile, it would be interesting to try to find a way to introduce a general concept of the proper time of the source, different from the universal time, applicable to aftershocks, foreshocks and mainshocks. Such an attempt was made in the works [23, 24]. The design of the "underground clock" was changed as follows.

In studying the global activity of strong earthquakes, the interval between two successive relatively weaker earthquakes was chosen as the basic unit of eigenvalue [23]. In other words, the counting of proper time was carried out in the most primitive way, using underground tremors of relatively small magnitude. More successful was the idea of studying the evolution of foreshocks and aftershocks using the foreshocks and aftershocks themselves as markers of the source's own time [24].

Let's look at this issue in more detail. To be able to evaluate the result of the transition to a new method of chronometry, let us consider the sequence of aftershocks. Let us number the aftershocks with natural numbers $N = 1, 2, 3, \ldots$ in the order of their occurrence. We will call $N$ the discrete proper time of the source. For each $N$ we determine the world time $t_N$ of the occurrence of the corresponding aftershock. Let us plot points $(N, t_N)$ on the coordinate plane $(C, t)$. Here the symbol $C$ (from the word *Chronos*) denotes the continuous proper time of the source. Now we approximate a finite set of points $(N, t_N)$ by a sufficiently smooth continuous function $t(C)$ and introduce an auxiliary function

$$\bar{g} = \frac{dt}{dC}. \tag{12}$$

The deactivation coefficient calculated using the above method is

$$\bar{\sigma} = \frac{d}{dC} \ln \langle \bar{g} \rangle . \tag{13}$$

Let us recall that formulas (1) and (8), as well as the auxiliary function $g(t) = 1/n(t)$, were chosen in such a way that, with $\sigma = \text{cjnst}$, the classical Omori law is strictly fulfilled when ordering events according to world time. If our underground clock works properly, then we have the right to expect that $\bar{g}(C)$ is a linear function, and $\bar{n}(C)$ is a hyperbolic function of time $t$, with

$$C(t) \propto \ln t, \tag{14}$$

and

$$\bar{\sigma} = \sigma \tag{15}$$

at $\sigma = \text{const}$. We can use the indicated properties of $\bar{g}$, $\bar{n}$, $C$, and $\bar{\sigma}$ as criteria for the rationality of our unusual choice of the design of the clock that counts the proper time of the source. Only experience with the use of new chronometry in earthquake research will show how effective it is in terms of allowing the discovery of previously unknown properties of seismicity. In this study we will limit ourselves to using (14) and (15) as tests for checking the running of the underground clock in one specific example.

Let us consider an event whose phase portrait is shown in Figure 2. It is convenient because the Omori epoch lasted a long time, more than three months. The deactivation coefficient was carefully measured and found to be $\sigma = 0{,}0004$. We want to compare $\sigma$ with the value of $\bar{\sigma}$ measured by the proper time of the source.

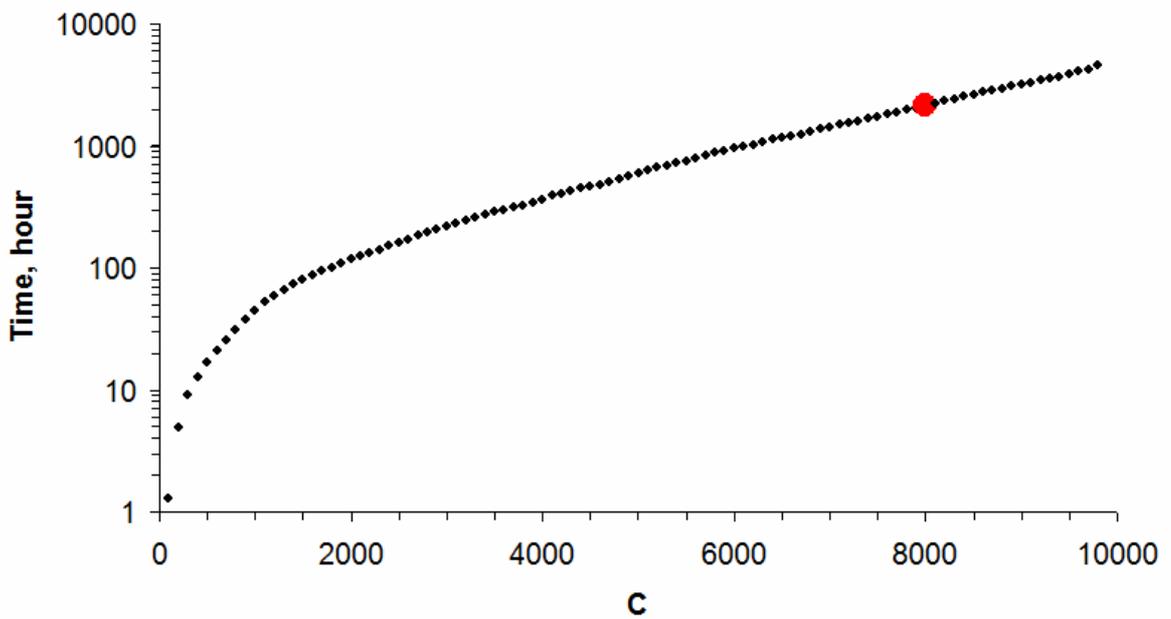

**Fig. 3**. Dependence of the world time of aftershock occurrence on the proper time of the earthquake source. The red dot shows the moment of the end of the Omori epoch.

Let us recall that 9867 aftershocks were registered in the 200 days after the main shock. Each aftershock received its own serial number $N$. For ease of graphical representation, we divided the aftershocks into consecutive clusters of 100 aftershocks each and determined the time of occurrence of the 51st aftershock in each cluster. Finally, we plotted the points found in this way on the coordinate plane ($C$, $t$) as shown in Figure 3. The figure clearly and convincingly demonstrates that throughout almost the entire evolution of the source, or more precisely, during 195 days out of 200, the discrete sequence of points belongs to a graph of type $\ln t \propto C$. Consequently, criterion (14) is fulfilled with the accuracy that can be required from experiments of this kind.

To test our underground clock using formulas (13), (15), we approximate the experimental points, starting from $N = 1501$ and ending with $N = 9867$, with a continuous smooth function

$$t(C) = a \exp[b(C-1)]. \qquad (16)$$

The calculations yield $a = 55.3$, $b = 0.0005$ with a very high coefficient of determination of $R^2 = 0.99$. Using formulas (12), (13) we find $\bar{\sigma} = 0.0005$, which differs from $\sigma$ by only 20%. We believe that this difference should not be given much importance. After all, $\bar{\sigma}$ and $\sigma$ were calculated using two essentially different methods. The fact that $\sigma$ was calculated based on data on aftershocks during the Omori epoch could also have had an effect, while to calculate $\bar{\sigma}$ we used a twice as long interval of the source evolution. However, we will conduct another independent check.

The universal time interval between successive aftershocks is $T_N = t_{N+1} - t_N$. The average waiting time for the next aftershock is shown in Figure 4 by a straight line

$$T(C) = 3.9 \exp[0.0004(C-1)]. \qquad (17)$$

The coefficient of determination is 0.98. It follows directly from this that $\bar{\sigma} = 0.0004$.

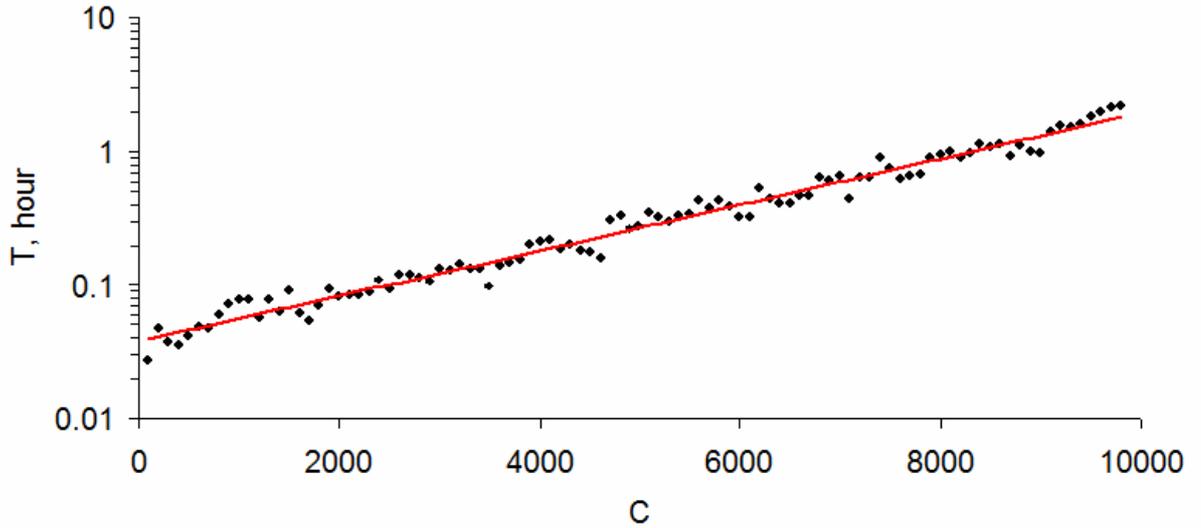

**Fig. 4**. Average waiting time for the next aftershock. The straight line approximates the points. Each point indicates the waiting time calculated for 100 consecutive aftershocks.

We will also not attach any special importance to the complete coincidence of $\bar{\sigma} = \sigma$. The main question in this whole matter is how useful is the measurement of proper time by our underground clocks for discovering new properties of earthquakes. For now, we can point out one non-trivial result of using the technique, the description and justification of which is given above. We are talking about the convergence of foreshocks and divergence of aftershocks, discovered in a statistical analysis of global seismicity [24].

In conclusion of the discussion, it is appropriate to present an equation for the frequency of aftershocks when ordered by the proper time of the source:

$$\frac{d\bar{n}}{dC} + \bar{\sigma}\bar{n} = 0. \qquad (18)$$

The solution of the equation has the form

$$\bar{n}(C) = \bar{n}_1 \exp\left[-\bar{\sigma}(C-1)\right] \quad (19)$$

for $\bar{\sigma} = \text{const}$, and

$$\bar{n}(C) = \bar{n}_1 \exp\left[-\int_1^C \bar{\sigma}(C')dC'\right] \quad (20)$$

in the general case. We found that the evolution of aftershocks is described by a linear differential equation. The aftershock frequency decreases exponentially with proper time. It should be noted that the evolution of the source in world time proceeds completely differently: the evolution equation (3) is nonlinear, and the attenuation of aftershocks occurs according to the hyperbolic law for $\sigma = \text{const}$. Let us emphasize that both equations, (3) and (18), describe the relaxation of the earthquake source holistically.

## 4. Conclusion

The idea of an elementary theory of aftershocks arose in an attempt to understand phenomenologically the dynamics of an earthquake source after the formation of a main rupture in the continuity of rocks. The phenomenological reduction, i.e. the transition from facts to essence, consisted of a radical transition from the natural view of the source as a complexly structured and complexly functioning structure to a simple idealized dynamic system.

The concept of deactivation of the earthquake source is central to elementary theory. In choosing the method for calculating the deactivation coefficient from aftershock frequency data, we deliberately used epoché (ἐποχή), refraining from any preliminary judgments about the law of aftershock evolution. As a result, it was possible to discover a two-stage relaxation mode of the source and establish that Omori's law is strictly fulfilled, but only at the first stage of evolution.

Within the framework of elementary theory, the idea of the proper time of the earthquake source arose, generally speaking, different from world time. We have proposed an original way of counting proper time. The proper time is logarithmically dependent on world time in the Omori epoch. A specific example shows that aftershock activity decreases exponentially with the proper time of the source.

*Acknowledgments*. We express our sincere gratitude to B.I. Klain and A.D. Zavyalov for fruitful discussions and support. We thank colleagues at the US Geological Survey for lending us their earthquake catalogs USGS/NEIC for use. The work was carried out within the framework of the planned tasks of the Ministry of Science and Higher Education of the Russian Federation to the Institute of Physics of the Earth of the Russian Academy of Sciences.